\documentclass[%
 reprint,
superscriptaddress,
 amsmath,amssymb,
 aps,
 pra,
]{revtex4-2}

\usepackage{graphicx}% Include figure files
\usepackage{dcolumn}% Align table columns on decimal point
\usepackage{bm}% bold math
\usepackage{xcolor}% textcolor
\usepackage[T2A]{fontenc}
\usepackage{float}

\usepackage{amsmath}
\usepackage{amssymb}
\usepackage{mathtools}
\usepackage{amsthm}
\usepackage{algorithm}
\usepackage{algpseudocode}
\usepackage{hyperref}
\usepackage{array} % For better column formatting
\usepackage{multirow} % For multirow cells
\usepackage{booktabs} % For professional-looking tables
\algnewcommand\algorithmicforeach{\textbf{for each}}
\algdef{S}[FOR]{ForEach}[1]{\algorithmicforeach\ #1\ \algorithmicdo}
\usepackage{enumitem}
\usepackage{subfigure}
\usepackage{physics}
\usepackage{placeins}

\DeclareMathOperator*{\argmax}{arg\,max}

\begin{document}

\preprint{AIP/123-QED}

\title[DQSP-MARL]{Distributed quantum architecture search using multi-agent reinforcement learning}% Force line breaks with \\

\author{Mikhail Sergeev}
 \email{Mikhail.Sergeev@skoltech.ru}
 \affiliation{Skolkovo Institute of Science and Technology, Moscow, Russian Federation}%Lines break automatically orf can be forced with \\
\author{Georgii Paradezhenko}
 \affiliation{Skolkovo Institute of Science and Technology, Moscow, Russian Federation}
\author{Daniil Rabinovich}
\affiliation{Skolkovo Institute of Science and Technology, Moscow, Russian Federation}%
\affiliation{The Russian Quantum Center, Moscow, Russian Federation}%
\affiliation{Moscow Institute of Physics and Technology, Dolgoprudny, Russian Federation}%
\author{Vladimir V. Palyulin}
 % \email{Mikhail.Sergeev@skoltech.ru}
 \affiliation{Skolkovo Institute of Science and Technology, Moscow, Russian Federation}

% \author{Mikhail Sergeev}
%  % \email{Mikhail.Sergeev@skoltech.ru}
%  \affiliation{Skolkovo Institute of Science and Technology, Bolshoy Boulevard 30, Moscow, 121205, Russia}
 
% \author{Vladimir V. Palyulin}
% \author{Placeholder\_name2}
% \affiliation{Skolkovo Institute of Science and Technology, Bolshoy Boulevard 30, Moscow, 121205, Russia}

\date{\today}% It is always \today, today,

\begin{abstract}

Quantum architecture search (QAS) automates the design of parameterized quantum circuits for variational quantum algorithms. The framework finds a well-suited problem-specific structure of a variational ansatz. Among possible implementations of QAS  the reinforcement learning (RL) stands out as one of the most promising. Current RL approaches are single-agent-based and show poor scalability with a number of qubits due to the increase of the action space dimension and the computational cost. We propose a novel multi-agent RL algorithm for QAS with each agent acting separately on its own block of a quantum circuit. This procedure allows to significantly accelerate the convergence of the RL-based QAS and reduce its computational cost. We benchmark the proposed algorithm on MaxCut problem on 3-regular graphs and on ground energy estimation for the Schwinger Hamiltonian. In addition, the proposed multi-agent approach naturally fits into the set-up of distributed quantum computing, favoring its implementation on modern intermediate scale quantum devices. 
\end{abstract}

\keywords{} %Use showkeys class option if keyword

\maketitle

\section{Introduction}

% Quantum computing is widely acknowledged as a potential tool to effectively solve various challenging computational problems with markedly lesser algorithmic complexity in comparison with classical computers \cite{shor1999polynomial}.

%Quantum computing is generally considered as a prominent tool for solving various challenging computational problems. An efficiently designed quantum algorithm can significantly reduce the algorithmic complexity in comparison to its classical counterpart for the same problem, e.g. Shor's algorithm \cite{shor1999polynomial}.

Variational quantum algorithms (VQA) are currently a standard approach for solving challenging computational problems on quantum computers \cite{cerezo2021variational}. Inspired by neural networks, they utilize parametrized quantum circuits (PQCs) with parameters updated by minimizing a given cost function. The execution of a PQC is carried out on a quantum processor, then components of a cost function are evaluated, followed by an update of the variational parameters on a classical co-processor~\cite{Peruzzo2014}. VQAs have been applied to diverse problems, including combinatorial optimization~\cite{farhi2014quantum,zhou2020quantum}, quantum chemistry~\cite{Cao19,tilly2022variational}, quantum error correction \cite{xu2019variational,cao2022quantum}, quantum machine learning (QML)~\cite{Biamonte2017,benedetti2019parameterized,Kardashin2025}, etc.
In practice, designing an efficient PQC (or ansatz) to solve a specific problem on a modern noisy intermediate-scale quantum (NISQ) device~\cite{Preskill2018quantumcomputingin} remains a non-trivial task.
Due to the limited fidelities of quantum gates, PQCs are required to be as shallow as possible, while still being expressive enough to find a solution~\cite{SJA2019}.
Problem-agnostic ansaetze with a fixed structure can be highly expressive such as, for instance, the hardware efficient ansatz~\cite{Kandala2017}. However, they suffer from trainability limitations such as barren plateaus~\cite{holmes2022connecting} and non-convex cost landscapes swamped by multiple local minima~\cite{Anschuetz2022}. One can deal with these limitations by using various optimization strategies including parameter initialization~\cite{Grant2019initialization} and concentrations \cite{akshay2021parameter}, quantum natural gradient~\cite{Wierichs2020}, layerwise training~\cite{akshay2022circuit, campos2021training}, etc. Alternatively, the performance of a VQA can be improved instead by looking for the optimal composition of a PQC tailored for a specific problem \cite{wiersema2020exploring, paradezhenko2025heuristic, rabinovich2022ion}. 

The search for an optimal problem-specific ansatz structure can be performed with the quantum architecture search (QAS) (for a review, see~\cite{martyniuk2024quantum}). 
The QAS framework consists of the exploration of a search space of PQCs. The search space specifies all possible architectures of PQCs under the restrictions imposed by a particular quantum hardware (supported gates, qubits' connectivity, etc.). Even reduced, this space is quite large and substantial computational resources to find the optimal circuit with standard optimization algorithms are required~\cite{he2022search}. The quantum circuit simulations are also quite costly for a large number of qubits. Thus, a search strategy for QAS should be as efficient as possible, minimizing the number of calls to a quantum simulator~\cite{patel2024curriculum}.

%{\color{red}Primarily, the immense variety of potential circuit architectures creates significant obstacles in efficiently exploring the search space. This necessitates a meticulous design of the exploration mechanisms and the configuration of the environment. Additionally, in the context of VQCs, the expansive action space significantly influences convergence due to the fact that minor differences in related actions, such as applying CNOT gates between adjacent qubits, can profoundly alter the overall state output by a VQC.
% Secondly, a high computational cost of quantum circuit simulations requires an algorithm for QAS to be as efficient as possible, accessing the system simulator the fewest number of times. Both of these challenges are feasible to overcome in the field of reinforcement learning, however, the training pipeline should be organized in such a way that the chosen algorithm should circumvent all major obstacles in the given problem. 
%Second, a high computational cost of quantum circuit simulations requires an algorithm for QAS to be as efficient as possible, accessing the system simulator the minimal number of times \cite{he2022search,patel2024curriculum}.}
% RL allows to overcome both of these challenges, but the training pipeline should be organized in such a way that a chosen algorithm circumvents the aforementioned obstacles in a given problem.

Various search strategies in QAS have been developed with most notable being meta-heuristics \cite{chivilikhin2020,Huang2022,stein2025incorporating}, adaptive methods~\cite{grimsley2019,Li2020}, differentiable quantum architecture search~\cite{zhang2022differentiable,sun2023differentiable}, random search~\cite{Du2022} and reinforcement learning (RL)~\cite{mckiernan2019automated,ostaszewski2021reinforcement,ye2021quantum,sogabe2022model, dai2024quantum,lockwood2021optimizing,kimura2022quantum,chen2023quantum,zhu2023quantum,promponas2024compiler,dai2024quantum,chen2024efficient,fodera2024reinforcement,Kundu2024,Rapp2025}. 
In the RL approach it is assumed that a PQC can be constructed as a sequence of actions performed by RL agent with a trainable policy. In particular, a RL algorithm based on the double deep-$Q$ network (DDQN)~\cite{vanhasselt2015deepreinforcementlearningdouble} was used in~\cite{ostaszewski2021reinforcement} for the automated ansatz design for finding the ground state of a lithium hydride (LiH) molecule in variational quantum eigensolver (VQE). In~\cite{Kundu2024}, the DDQN technique was applied to find a problem-inspired ansatz for the variational quantum state diagonalization algorithm~\cite{LaRose2019}. The proximal policy optimization algorithm~\cite{schulman2017proximal} was exploited in~\cite{mckiernan2019automated,fodera2024reinforcement} to generate PQCs well-suited for solving specific combinatorial optimization problems.
The planning algorithm based on the quantum observable Markov decision process (MDP)~\cite{Barry2014} was proposed in~\cite{kimura2022quantum}, where it was applied to circuit design for the problems of state preparation and energy minimization in VQE. In QML, the quantum circuit design for classification problems was addressed by DDQN in~\cite{dai2024quantum} and model-based MuZero~\cite{Schrittwieser2020} algorithm in~\cite{Rapp2025}.
However, most of these studies do not address the scalability of RL algorithms for QAS directly which limits their possible applications. 

%In the works \cite{ostaszewski2021reinforcement, ye2021quantum, sogabe2022model, dai2024quantum}, various approaches based on off-policy RL algorithms are presented and tested in different scenarios. In \cite{kimura2022quantum}, the authors adopt the Markov decision process properties to effectively update quantum circuit structure. Quantum RL application for QAS is shown in \cite{chen2023quantum}. On-policy RL algorithm is presented in \cite{zhu2023quantum}. 
% Circuit optimization routine for RL problems can be found in \cite{kolle2024architectural}, where VQAs are used as the agents in MARL scenario, with evolutionary algorithm used to optimize agents' circuit structure. 
%RL approaches to QAS in quantum compilation are shown in \cite{promponas2024compiler, chen2024efficient}. Concerning the importance of optimization algorithm for PQC parameters, in \cite{lockwood2021optimizing}, using RL for parameters optimization does not offer a significant improvement in comparison to conventional gradient descent. In \cite{fodera2024reinforcement}, the version of Proximal Policy Optimization algorithm \cite{schulman2017proximal} is used to optimize circuits for graph partitioning problems (e.g. MaxCut \cite{commander2009maximum}). Model-based algorithm MuZero is used for QAS for QML problems in \cite{Rapp2025} 

In this work, we address the scalability problem by introducing a novel QAS approach based on multi-agent reinforcement learning (MARL) \cite{ning2024survey}. The core of our method is the multi-agent version of deep-$Q$ learning known as QMIX~\cite{rashid2020monotonic}, proven to be effective for collaborative multi-agent tasks. Our method is inspired by the recently proposed distributed QAS framework~\cite{situ2024distributed}, which allows to automatically design PQCs for interconnected quantum processing units in distributed quantum computing (DQC) (for a review, see~\cite{caleffi2024distributed}).
Specifically, we partition a multi-qubit system into small subsystems, interconnected through two-qubit gates between nearest neighbor qubits. Each subsystem is supervised by its own RL agent, while all agents collaborate to maximize their common reward. Thus, our multi-agent approach can be naturally adopted to the framework of DQC.

The proposed algorithm is benchmarked on the Hamiltonian minimization problems for the Max-Cut combinatorial optimization~\cite{zhou2020quantum} and the Schwinger model~\cite{Kokail2019} for the system sizes up to 12 qubits. For both problems, we apply our approach to design problem-specific PQCs and compare them to the standard circuits in QAOA~\cite{farhi2014quantum} and VQE~\cite{Kandala2017}, respectively. We show that the designed circuits have several advantages over the standard ones, including the reduced number of two-qubit CNOT gates and variational parameters, which favors their implementation on NISQ devices. Moreover, we examine the effect of multiple agents in the RL-based QAS and demonstrate that, in comparison to a single-agent RL, the MARL approach allows to accelerate the convergence of the RL training process and reduce the computational cost. As the result, the MARL alleviates the exploration of large action spaces in QAS similarly to its recent applications demonstrated for various problems (see, e.g.,~\cite{Zhang2021,vigon2023effective,heik2024study}).

The paper is organized as follows.
In Section ~\ref{subsect:RL}, we provide a brief introduction to RL. Then, in Section ~\ref{subsect:QMIX}, we explain the core idea of the QMIX method for training multi-agent systems. In Section ~\ref{subsect:MARL-QAS}, we introduce a MARL approach for searching problem-specific quantum circuits in QAS. Section ~\ref{subsect:setup} describes the setup of our simulations. Sections ~\ref{subsect:combinatorial} and \ref{subsect:Schwinger} show the numerical results, where we benchmark the proposed MARL-QAS approach on the Max-Cut and Schwinger problem Hamiltonians. In Section ~\ref{sect:conclusion} the results are concluded and discussed.

\section{Methods}
\label{sect:methods}
% In this section, a brief introduction to variational quantum circuits will be given. Also, an explanation of reinforcement learning and multi-agent reinforcement learning will be provided as well. Finally, a description of the QMIX MARL algorithm used in this paper will be presented.

% In this section, we provide a brief introduction to variational quantum circuits followed by a detailed explanation of the RL and multi-agent RL. Finally, we describe the QMIX MARL algorithm. 
\label{chap:MARL}

% сюда можно вставить дистрибудет часть

\subsection{Reinforcement learning}\label{subsect:RL}

\begin{figure}[ht]
    \centering
    \includegraphics[width=\linewidth]{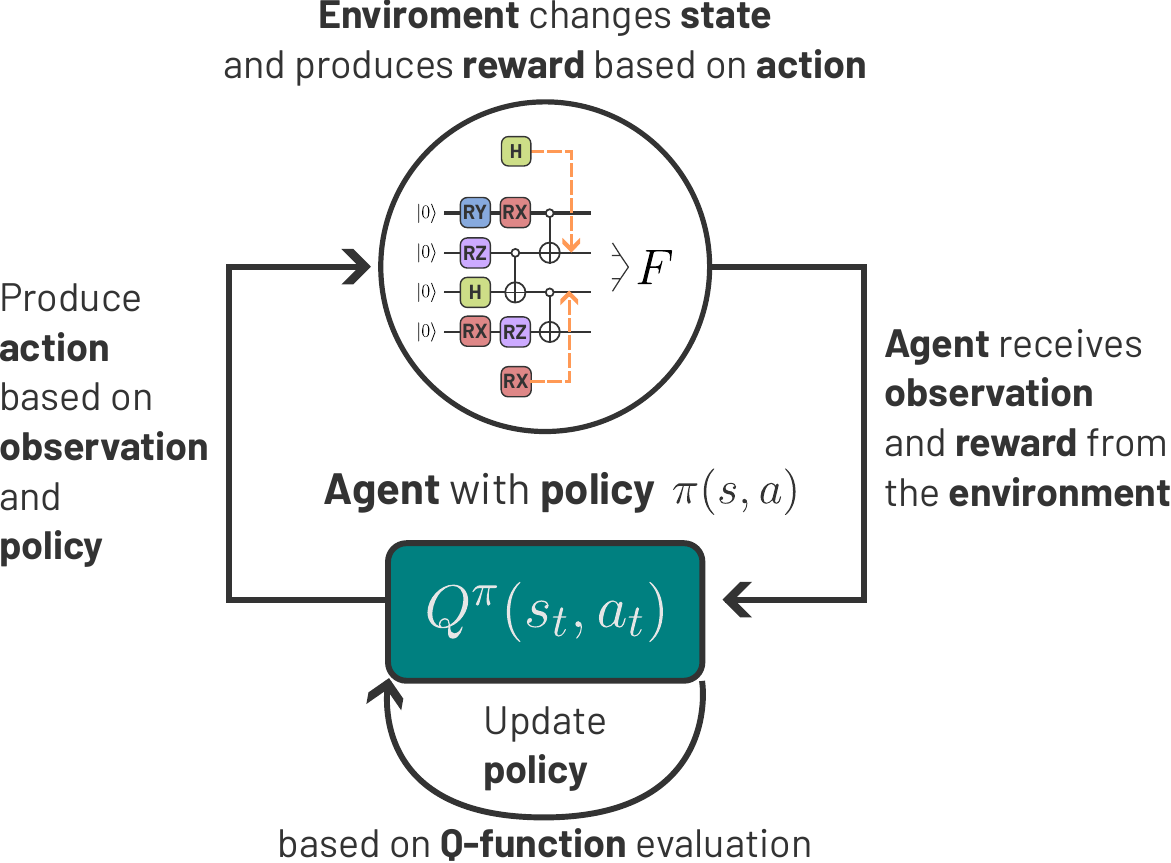}
    \caption{Sketch of a training loop in basic RL. An agent receives the observation vector $o$ and reward $r$ from the environment. Based on this information, the agent applies an action on the environment and changes its state. The agent is trained to maximize the reward by updating its action policy $\pi$ according to a certain rule. In $Q$-learning algorithms, the agent updates its $Q^{\pi}(s,a)$ function given by Eq.~\eqref{Q-function}.}
    \label{fig:BasicRL}
\end{figure}

Reinforcement learning is a class of machine learning algorithms that learn by training an agent to choose an optimal sequence of actions through interaction with a dynamic environment to maximize a reward signal~\cite{Sutton2018}. The training of a basic RL model is schematically shown in Fig.~\ref{fig:BasicRL}. 
The set of states $s \in S$ forms an environment. The agent exercises an action $a \in A$ on the environment and, in turn, the latter returns a reward $r$, which serves as a learning signal. The corresponding reward function $r \equiv r(s,a): S \times A \to \mathbb{R}$ is problem-dependent. During the training phase, the RL agent learns an action policy $\pi \equiv \pi(s,a): S \times A \to [0,1]$, which describes the probability of taking an action $a$ when the environment is in a state $s$. The aim is to maximize the cumulative reward along a possible trajectory,
\begin{equation}\label{reward-cumulative}
    R = \sum_{t=0}^\infty\gamma^{t} r_t,
\end{equation}
where $t$ is the step number, $\gamma \in [0,1)$ is the reward discount factor introduced for preventing infinite state transition loops. During the execution phase, the agent applies actions according to the learned optimal policy $\pi^*$, but does not receive any feedback from the environment. 

A basic RL algorithm can be modeled as a Markov decision process (MDP), which guarantees that the training converges to an optimal solution. The most prominent example is the $Q$-learning algorithm~\cite{watkins1992q}. This off-policy RL algorithm is based on the concept of an action-value function $Q$, also referred to as $Q$-value:
\begin{equation}\label{Q-function}
    Q^{\pi}(s,a):=\mathbb{E}_{\mathcal{T}\sim\pi | s_0=s,a_0=a} \left[ R \right],
\end{equation}
where the expectation value of \eqref{reward-cumulative} is taken over all possible trajectories $\mathcal{T}:=\{s_0,a_0,s_1,a_1, \ldots \}$ of the episode starting from the state $s$, taking the action $a$, and thereafter following the policy $\pi$. The training process in the $Q$-learning algorithm is governed by the update formula
\begin{eqnarray}
    Q(s_t,a_t) & \leftarrow & (1 - \alpha)\,  Q(s_t,a_t) \nonumber \\ 
    & + & \alpha \left[ r_{t+1} + \gamma \max_{a} Q(s_{t+1},a) \right],
    \label{Q-update}
\end{eqnarray}
where $\alpha$ is the learning rate. After a sufficient number of updates~\eqref{Q-update}, the algorithm converges in such a way that the highest $Q$-values are assigned to the optimal state-action pairs.
Based on these stored $Q$-values, the optimal policy is determined as $\pi^*(s) =\argmax_{a}Q(s,a)$.

\begin{figure*}[t]
    \centering
    \includegraphics[width=\linewidth]{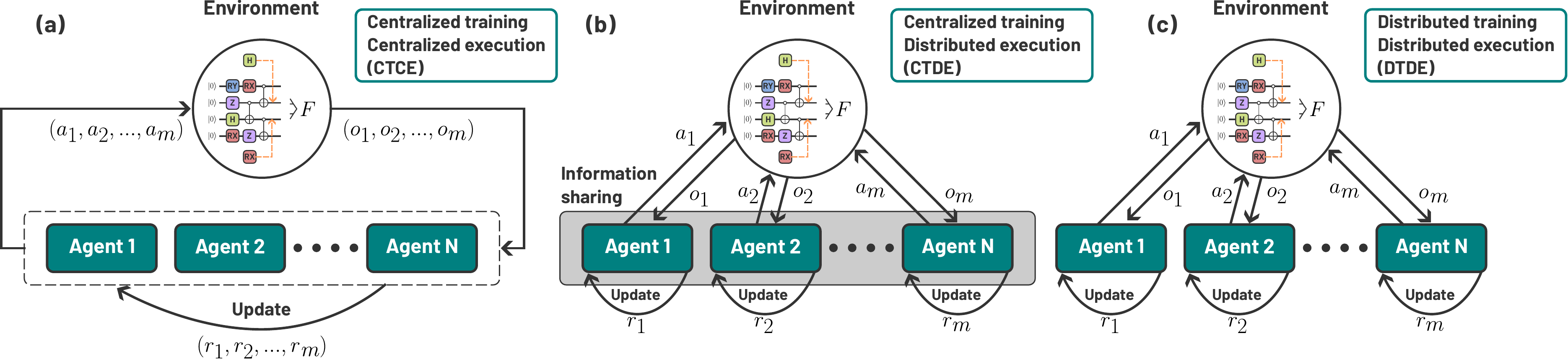}
    \caption{Forms of the agent-environment interaction in multi-agent reinforcement learning. In all cases, multiple agents receive an array of corresponding observations, $\textbf{o}=(o_1,o_2,...,o_n)$, and apply actions represented as an array $\textbf{a}=(a_1,a_2,...,a_n)$. The agents receive their rewards $r_1,r_2,...,r_n$ from the environment. These rewards can be joint for cooperative problems or separate otherwise. (a) Centralized training, centralized execution: agents can receive all information about other agents during the training and execution phases. (b) Centralized training, distributed execution: agents can share information during the training phase, but during the execution they act independently. (c): Distributed training, distributed execution: agents do not share any information during both the training and the execution phases.}
    \label{fig:MARL_taxonomy}
\end{figure*}

In practice, it is costly to store $Q$-values for all state-action pairs, especially for large RL models. Instead of a tabular storage, the values of $Q$ can be evaluated using a neural network called a deep $Q$ network (DQN)~\cite{mnih2013playing}. In DQN, the $Q$-function~\eqref{Q-function} is evaluated as an output $Q \equiv Q(s,a; \bm{w})$ of a neural network with weights $\bm{w}$ trained by minimizing the loss function
\begin{eqnarray}
    \mathcal{L}(\bm{w}) & = & \sum^b_{k=1}[(y^{\rm DQN}_k -Q(s,a;\bm{w})^2], \label{DQN-loss} \\
    y^{\rm DQN} & = & r + \gamma\max_{a'}Q(s',a';\bm{w}^-). \label{DQN-target}
\end{eqnarray} 
For calculating the loss~\eqref{DQN-loss}, a batch of $b$ state transitions is sampled from a replay buffer, which stores the transition tuples $\langle s,a,r,s'\rangle$, where the state $s'$ is observed after taking the action $a$ in the state $s$ and receiving the reward $r$. In Eq.~\eqref{DQN-target}, $\bm{w}^-$ are the parameters of a target network that is the same network as $Q(s,a;\bm{w})$, but its parameters are periodically copied from $\bm{w}$ and kept constant for a number of training steps.

While originally introduced for a single-agent-based RL models the DQN approach can be extended to the multi-agent reinforcement learning (MARL) (for a review, see~\cite{ning2024survey}). In MARL, the interaction between multiple RL agents and their common environment can be implemented in three different ways (see Fig.~\ref{fig:MARL_taxonomy}): 
\begin{itemize}[leftmargin=*]
    \item Centralized Training, Centralized Execution (CTCE);
    \item Centralized Training, Distributed Execution (CTDE);
    \item Distributed Training, Distributed Execution (DTDE).
\end{itemize}
The difference between these approaches depends mainly on the way the information is exchanged between agents during the training or execution phases. 
%In this paper, we focus only on the training phase of MARL. 
Since QAS is a fully cooperative multi-agent problem, where agents share the common goal and reward, we consider the training to be centralized. The execution phase can be chosen to be centralized (CTCE) allowing to consider a multi-agent problem as its single-agent counterpart. Alternatively, being distributed (CTDE), it is well-suited for the distributed computation.

Similarly to a single-agent-based RL, one can describe the training of a fully cooperative MARL system as a partially observed MDP $G = \langle S,A,P,r,\mathcal{O},O,m,\gamma\rangle$ \cite{oliehoek2016}. Here, $s \in S$ is the state of the environment, the agents labeled by the index $i \in \{ 1, \ldots, m\}$ apply actions $a_i \in A$, while $\bm{a} \in \bm{A} \equiv A^m$ forms their joint action on the environment. Next, $P(s'|s,\bm{a}):S\times\bm{A}\times S \rightarrow[0,1]$ is the state transition function determined by the environment dynamics, $r(s,\bm{a}):S\times \bm{A} \rightarrow \mathbb{R}$ is the common reward function for all agents, and $\gamma\in[0,1)$ is the reward discount factor. 
Finally, for a partially observed MDP, each agent draws individual observations $o \in \mathcal{O}$ based on the observation function $O(s,a): S \times A \to \mathcal{O}$, and it has its own action-observation history $\tau^i \in T \equiv (\mathcal{O} \times A)^*$. Here, the asterisk means that the sequence of previous action-observations pairs can be extended to an arbitrary number of steps, $\tau^i = \{ o_0^i, a_0^i, o_1^i, a_1^i, \ldots\}$. 

In this formulation, each agent has its own action policy $\pi_i(a^i|\tau^i): T \times A \rightarrow [0,1]$, while the action-value function for the joint policy $\pi$ reads
\begin{equation}\label{Q-fun-MARL}
    Q^{\pi}(s_t,\bm{a}_t) = \mathbb{E}_{\mathcal{T} \sim \pi|s_0 = s_{t}, \bm{a}_0 = \bm{a}_{t+1}}[R_t],
\end{equation}
where $R_t = \sum^\infty_{t'=0} \gamma^{t'}\, r_{t+t'}$ is the cumulative reward, and its expectation value is taken over all possible trajectories $\mathcal{T}:=\{s_0,\bm{a}_0,s_1,\bm{a}_1, \ldots \}$ of the episode starting from the state $s_t$, taking the joint action $\bm{a}_t$, and thereafter following the joint policy $\pi$.

\begin{figure*}[ht]
    \centering
    \includegraphics[width=0.9\linewidth]{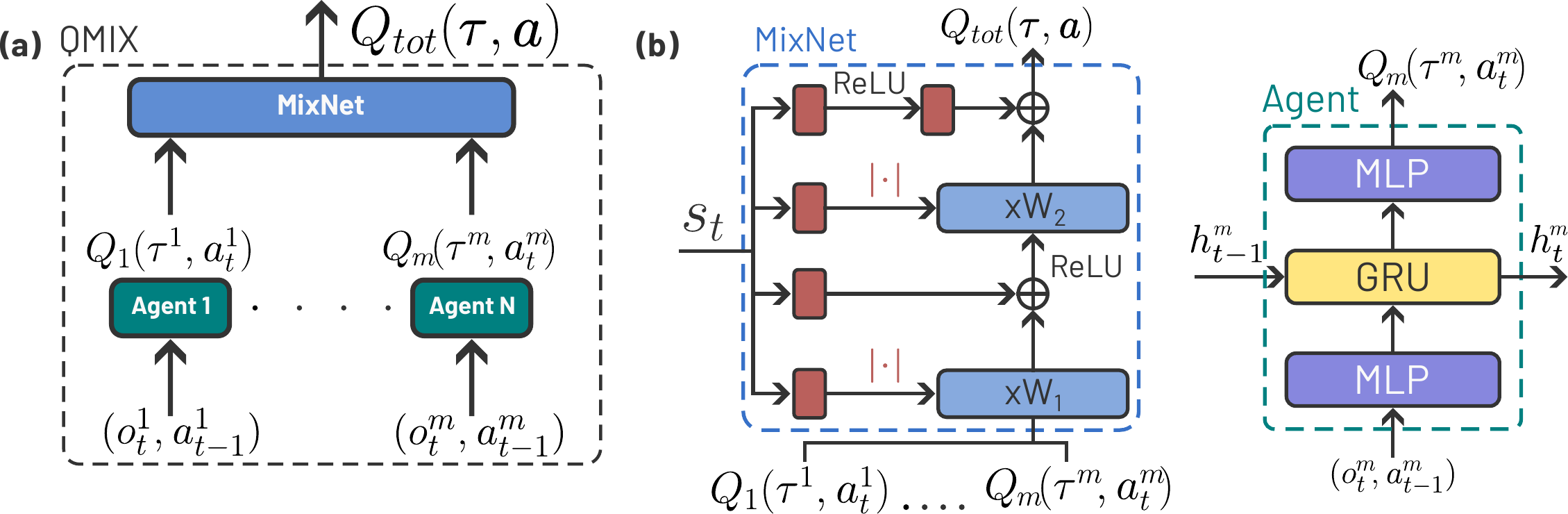}
    \caption{(a) General structure of the QMIX network. This network consists of multiple agents networks connected to the single mixing network (MixNet). Each agent network receives the action-observation pair $(o^i_t, a_{t-1}^i)$ as an input and outputs the evaluated Q-function $Q_i(\tau^i,a^i_t)$, while the MixNet combines all these Q-values into a single total Q-function $Q_{tot}(\bm{\tau},\bm{a}_t)$ for the joint action $\bm{a}_t$. (b) Structures of the MixNet and agent network. The MixNet consists of two linear layers with a ReLU activation function between them. The weights and biases of MixNet are produced by separate hypernetworks with trainable weights that use the total state $s_t$ of the environment as an input. The agent network consists of two linear layers connected through a single gated recurrent unit (GRU).
} 
    \label{fig:QMIX}
\end{figure*}

% {\color{red}Поменять порядок: начать с QMIX и объяснить Рис.3}
\subsection{QMIX algorithm}\label{subsect:QMIX}

For training a multi-agent system, we use the QMIX method~\cite{rashid2020monotonic}, which is one of the most prominent algorithms for CTDE in the value-based MARL. In QMIX, the total action-value function~\eqref{Q-fun-MARL} is calculated as
\begin{equation}\label{Q-tot}
    Q_{\rm tot} = f_s(Q_1(\tau_t^1,a^1_t),...,Q_m(\tau_t^m,a^m_t))
    \equiv Q^{\pi}(s_t,\bm{a}_t).
\end{equation}
It is conditioned by the global state $s$ and composed of action-value functions $Q_i$ for each agent $i$. The function \eqref{Q-tot} is used only during the centralized training phase. In practice,  this function is evaluated based on a specific neural network described below such that $Q_{\rm tot} \equiv Q_{\rm tot}(...,\bm{w})$, where $\bm{w}$ are the trainable weights.
The existence of $Q_{\rm tot}$ in cooperative MARL significantly improves the convergence of the training process in comparison to a completely independent training, where agents do not cooperate~\cite{rashid2020monotonic}. $Q_{\rm tot}$ should be monotonic with respect to each $Q_i$:
\begin{equation}\label{Q-monot}
    \frac{\partial Q_{\rm tot}}{\partial Q_i} \geq0, \quad \forall \, i=1,\ldots,m.
\end{equation}
In QMIX, $Q_{\rm tot}$ is represented by a mixing neural network (MixNet), which combines all $Q_i$ evaluated by neural networks of every RL agent in a non-linear fashion (see Fig.~\ref{fig:QMIX}a). As a result, QMIX can describe a complex inter-agent cooperation, while retaining a simple structure that scales linearly with the total number of agents. 

The MixNet is a feed-forward neural network, which receives the values of Q-functions from each agent, $Q_1(\tau^1,a^1_t),\ldots,Q_m(\tau^m,a^m_t)$, and outputs the Q-function for the joint state-action space, $Q_{\rm tot}(\bm{\tau},\bm{a}_t)$, evaluated by two linear layers (Fig.~\ref{fig:QMIX}b). The weights and biases of these two linear layers are produced by separate hypernetworks that use the total state $s$ of the environment as an input. However, the structure of these hypernetworks is different. For producing the MixNet weights, depicted as $\bm{W}_1$ and $\bm{W}_2$ in Fig.~\ref{fig:QMIX}b, the hypernetworks consist of a single linear layer followed by the absolute activation function. The latter enforces the monotonicity constraint~\eqref{Q-monot} by keeping the weights $\bm{W}_1$ and $\bm{W}_2$ non-negative.
The first bias is produced by a hypernetwork with a single linear layer without the absolute activation function, because the biases in MixNet are not restricted to be non-negative. The second bias is produced by a hypernetwork, which consists of two linear layers with a rectified linear unit (ReLU) activation function between them to incorporate non-linearity. The weights of all hypernetworks are trainable and are assumed to be a part of $\bm{w}$. The dimensions of two linear layers in MixNet are the same and are considered as a hyperparameter.

The agent networks are structured as follows (Fig.~\ref{fig:QMIX}b). For each agent $a$, its network takes the action-observation pair $(o^i_t, a_{t-1}^i)$ as an input and outputs the evaluated Q-function $Q_i(\tau^i,a^i_t)$. The agent network is composed of two linear layers and a single gated recurrent unit (GRU) cell between them, where $h$ is the hidden recurrent channel~\cite{cho2014properties}. The trainable weights of these layers are incorporated in $\bm{w}$. The dimensions of linear layers and a GRU cell are considered as hyperparameters.

The weights $\bm{w}$ of the QMIX neural network described above are dynamically updated at each step of the training process by minimizing the loss function similar to that of DQN (see Eqs.~\eqref{DQN-loss} and \eqref{DQN-target}):
\begin{eqnarray} 
    \mathcal{L}(\bm{w}) 
    & = & \sum^b_{k=1}\bigl[(y^{\rm tot}_k - Q_{\rm tot}(\bm{\tau},\bm{a}, s, \bm{w}) \bigr]^2, \label{loss-QMIX} \\
    y^{\rm tot} & = & r + \gamma \max_{\bm{a}'}Q_{\rm tot}(\bm{\tau}',\bm{a}', s',\bm{w}^-). \label{y-tot-QMIX}
\end{eqnarray}
Similar to DQN, QMIX samples state transitions $\langle s,\bm{a},\bm{\tau},r,s',\bm{\tau}' \rangle$ from a replay buffer using the QMIX target network with the weights $\bm{w}^-$. 
The replay buffer is collected by means of the $\epsilon$-greedy policy~\cite{Sutton2018}. In this context, $\epsilon$ serves as a hyperparameter and decreases linearly over a given number of steps.

\subsection{Multi-agent approach for QAS}\label{subsect:MARL-QAS}

% Distributed multi-agent approach can reduce the computational load, since the size of the matrix state is $(2^n,2^n)$, where $n$ is the number of qubits.
% By partitioning the circuit, we gain a computational speedup (as well as a smaller quantum device size) at the cost of losing  information about the similarity between the total state produced by the agent circuits and the target state vector and the need to carefully design the environment to distribute entanglement between agents, as shown in Fig.\ref{fig:distr}. 
% The algorithm and environment proposed in this paper serves as an example of implementation of an RL-based QAS strategy for distributed VQA.

In this work, we focus on QAS~\cite{martyniuk2024quantum} for the automated design of PQCs tailored for minimizing specific problem Hamiltonians in VQAs. Solving many real-world problems by VQAs requires a large number of qubits. However, this number is very limited by the capabilities of modern quantum devices, which support from tens to hundreds of noisy qubits~\cite{guerreschi2019qaoa}. For scaling quantum algorithms, a distributed quantum computing (DQC) approach has been developed (for a review, see~\cite{caleffi2024distributed}). The DQC relies on connecting multiple small quantum processors into a single cluster, where individual devices can exchange information (classical and/or quantum) using an interconnect \cite{main2025distributed}.
Recently, a distributed QAS framework for designing distributed quantum circuits for interconnected quantum processors has been proposed in~\cite{situ2024distributed}. This has motivated us to develop a novel QAS approach based on the MARL that naturally fits into the DQC paradigm. 

\begin{figure}[ht!]
    \centering
    \includegraphics[width=1\linewidth]{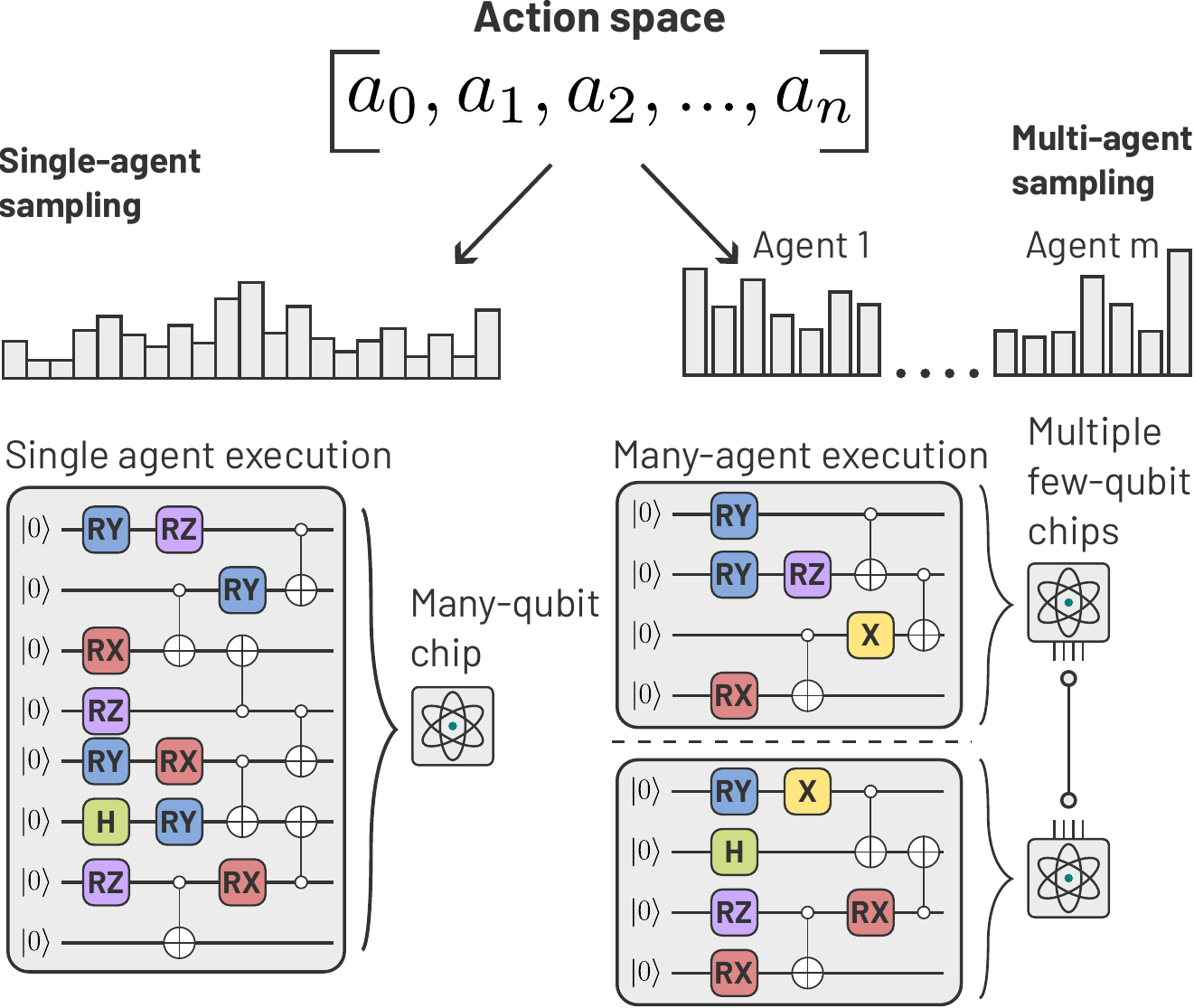}
    \caption{Sketch of the MARL-QAS algorithm, which allows to distribute the action space between multiple RL agents. 
    The agent-based execution model provides a potential for using the method in distributed quantum computing. 
    }
    \label{fig:distr}
\end{figure}

Following the DQC framework, we partition a quantum system into small equally-sized subsystems (see Fig.~\ref{fig:distr}). Each subsystem is supervised by its own RL agent. During the training phase of MARL, the agents learn how to design their own parts of the whole circuit, interconnected by two-qubit quantum gates, cooperating to maximize their common reward. Since the training is centralized, it requires the communication between individual quantum devices to calculate the reward and return it to each agent. However, during the execution phase of MARL, the agents can apply their optimal policies separately on each device to design subcircuits, which are further interconnected into the circuit with an optimal structure. In addition to the integrated distribution mechanism, the MARL approach for QAS partitions the action space between multiple agents, which allows to improve the exploration process~\cite{Zhang2021,vigon2023effective,heik2024study}.

\begin{figure*}
    \centering
    \includegraphics[width=\linewidth]{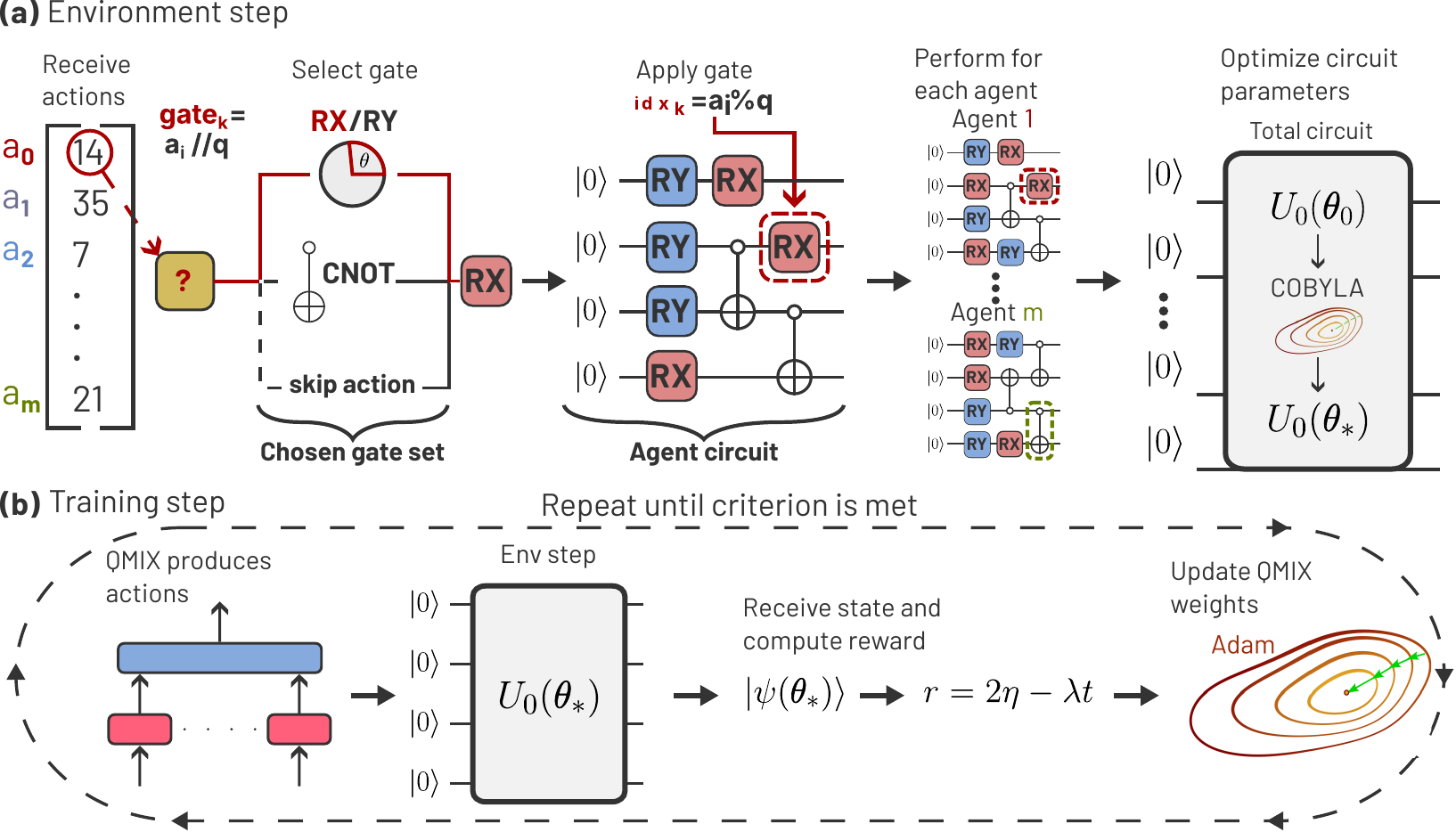}
    \caption{(a): Agent action-to-unitary transition. First, the gate is selected by pointwise division of $a$ by number of agent qubits $q$. The matching qubit index of gate application is selected by calculating the remainder of the same division. This routine is performed for each agent. (b): MARL training pipeline. RL algorithm produces a quantum circuit. Afterwards, the output circuit is assessed by computing the total statevector (or Hamiltonian expected value) and then by calculating the corresponding reward metric (fidelity $F$/approxim. ratio). According to the reward signal, the RL networks' weights are updated, for both mixing and agent networks.}
    \label{fig:Action_decomp}
\end{figure*}

In our MARL-QAS algorithm (see Fig.~\ref{fig:Action_decomp}b), during the interaction with the environment, multiple RL agents construct a PQC by choosing new quantum gates from the following set:
\begin{equation}\label{G-set-reduced}
    \mathcal{G}=\{R^{x}_j(\theta), R^{y}_j(\theta), \mathrm{CNOT}_{j,j- 1},\mathrm{CNOT}_{j,j+ 1}\},
\end{equation}
where $j \in \{1,\ldots, n\}$ is the qubit index, $n$ is the number of qubits, $R^{\alpha}_j(\theta) = e^{-i\,\sigma^\alpha_j \theta}$ is the single-qubit rotational gate and $\sigma^\alpha$ are the Pauli matrices ($\alpha = x,y$).
In this setting, we assume that the controlled NOT gates $\mathrm{CNOT}_{j,j \pm 1}$ connect only adjacent qubits and the $n$-th and the 1-st qubit are linked with periodic boundary conditions (PBC).

Each agent supervises its own part of the whole circuit, partitioned into equally-sized sub-circuits. For each agent $i$, its action space $A$ is described as a set of positive integers encoding a specific gate $\mathcal{U} \in \mathcal{G}$. A separate action token indicates that an RL agent skips its action. For decoding of a given integer action token $a^i$ into a specific quantum gate applied at a certain part of the circuit, we divide it by the number of qubits per agent $q$: $a^i = kq + l$. The integer part $k$ of this division provides an index to select a quantum gate from~\eqref{G-set-reduced}, while the remainder indicates a $(l+1)$-th qubit to apply the chosen gate. If a two-qubit gate is selected ($\mathrm{CNOT}_{j,j+1}$ or flipped $\mathrm{CNOT}_{j,j-1}$), then it is applied to the pair of qubits with indices $l$ and $l\pm 1$, respectively.

%This algorithm is shown at Fig.~\ref{fig:Action_decomp}. By using the proposed action space encoding, we get the adaptable mechanism to incorporate any unitary set $\mathcal{G}$ avoiding manual manipulations on the environment.

% In addition, the entanglement distribution can be composed using various approaches, each of them with its own set of advantages and drawbacks. The first approach in this paper involves extending the action space for each agent to enable inter-agent entanglement distribution. In this case, action space is increased by 4 possible actions (CNOT and flipped CNOT for ($k*m$,$k*m+1$) and ($k*m-1$,$k*m$) qubit pairs where $k$ is the agent index, $m$ is the number of qubits in each agent. The other two CNOT possible action probabilities are masked for those agents. According to that approach, an arbitrary entanglement distribution, without encumbrance on agents action space, can be prepared. On top of that, neighboring CNOTs are easier to implement than long-range CNOTs due to limited connectivity in different qubit platforms.

Once new gates are added to the circuit, we optimize all its variational parameters $\bm{\theta}$ by means of the COBYLA optimizer~\cite{powell1994direct}, which is a derivative-free method widely used for optimizing PQCs (see, e.g., \cite{Arrasmith2021effectofbarren}). As an initial guess, we use random parameters sampled from the uniform distribution on the parameter space. The optimized circuit provides a final state of the environment obtained in response of the joint action of RL agents on the current iteration of the RL training process. 

After their interaction with the environment, the RL agents receive the common reward
\begin{equation}\label{reward}
    r = 2 \eta - \rho t,
\end{equation}
where $\eta$ is the problem-specific cost evaluated based on the quantum state $\ket{\psi(\bm{\theta}_*)}$, which is prepared by the optimized PQC; $\rho$ is the circuit length penalizing term, and $t$ is the environment step. Additionally, the reward function can be further modified, e.g. if further penalization is needed for the amount of two-qubit gates in the circuit, an additional term into the expression \eqref{reward} could be added as well.
In this paper, we consider Hamiltonian minimization problems, where $\eta$ is given by the normalized approximation ratio, 
\begin{equation}\label{approx-ratio}
    \eta = \frac{\lambda_{\max} - E(\bm{\theta}_*)}{\lambda_{\max} - \lambda_0},
\end{equation} 
where $E(\bm{\theta}_*)$ is the optimized energy in VQA,
\begin{equation}\label{cost}
    E(\bm{\theta}_*) = 
    \min_{\bm{\theta}} \bra{\psi(\bm{\theta})} H \ket{\psi(\bm{\theta})}.
\end{equation}
In Eqs.~\eqref{approx-ratio}-\eqref{cost}, $H$ is the problem Hamiltonian, $\lambda_0$ and $\lambda_{\max}$ are the energies of the ground state and highest excited state of $H$. 
Once the reward is calculated, the environment step is considered to be complete. It is important to mention that $\lambda_{\rm max},\lambda_0$ are not necessarily have to be calculated explicitly (therefore, estimations are satisfactory for this formula as well) since the main principle of RL algorithms is the reward maximization.

Each training step of the QMIX-based QAS approach is organized as follows (see Fig.~\ref{fig:Action_decomp}a):
\begin{enumerate}
    \item The set of actions for RL agents is $\epsilon$-greedy taken from the target QMIX network (with the weights $\bm{w}^-$) based on the joint action-observation pairs from the previous environment step.
    \item The sampled actions are decoded into quantum gates applied to the circuit as explained in Fig. \ref{fig:Action_decomp}b.
    \item The variational parameters of the designed circuit are optimized for a specific problem using the COBYLA optimizer, where an initial guess is sampled randomly.
    \item The common reward~\eqref{reward} is calculated based on the optimized variational ansatz $\ket{\psi(\bm{\theta}_*)}$.
    \item The completed environment step (observations, actions, reward) is stored into the replay buffer. 
    \item The environment episode is considered to be finished, if either the maximum allowed number of environment steps is exceeded or a specified reward threshold is reached. 
    % After that, a new environment episode with a blank circuit starts.
    \item Based on the updated replay buffer, the current network updates its weights by minimization of loss function \eqref{loss-QMIX} for a batch of sampled states from the buffer. After a specified number of training steps (which serves as a hyperparameter), the weights of the target network are replaced by the current network's weights.
\end{enumerate}

% entanglement phase позволяет сэкономить на числе агентов 
\section{Numerical results}
\label{sect:results}

\subsection{Simulation setup}\label{subsect:setup}

%The training pipeline of the QMIX algorithm exhibits similarities to that of the DQN, thereby retaining certain hyperparameters commonly associated with the latter. Both QMIX-specific hyperparameters and these common parameters were consistently selected for addressing the MaxCut and energy estimation challenges. According to the QMIX publication, the architecture of the agent and mixing networks is delineated as follows: each agent's network is comprised of twin linear layers integrated with a hidden GRU consisting of 64 dimensions, while the mixing network is defined by the presence of a solitary hidden layer also containing 64 dimensions. As previously discussed, the training process incorporates an $\epsilon$-greedy exploration approach to accumulate state transitions. Here, $\epsilon$ is initialized at 1 and undergoes a decay over the course of 5000 steps. The recent experiences collected from 2000 episodes are maintained within a replay buffer, from which uniformly sampled batches of size 32 are utilized for training purposes. Moreover, every 400 episodes, the target network undergoes an update. To bolster learning efficiency, parameter sharing is implemented among agents, and the agent identifier is integrated into each observation vector to introduce differentiation. The discount factor $\gamma$ is maintained at a value of 0.99. To perform network weight updates, the Adam gradient optimizer \cite{kingma2014adam} is employed, with the learning rate $\alpha$ being established at $10^{-4}$.

We benchmark the proposed MARL-QAS approach on two illustrative examples of possible applications of VQAs: the Max-Cut combinatorial optimization problem~\cite{commander2009maximum} and the ground state search for the Schwinger model Hamiltonian~\cite{Kokail2019}.
For both examples, we apply our approach to design well-suited problem-specific PQCs and compare their performance to the standard QAOA~\cite{farhi2014quantum} and VQE~\cite{Peruzzo2014} algorithms.
Our simulations are carried out systematically for the system sizes $n$ of 4, 6, 8, 10 and 12 qubits. The QMIX training pipeline is configured as follows. The dimensions of linear layers in both MixNet and agents networks are set to 64. The $\epsilon$-greedy policy is implemented such that we set $\epsilon=1$ initially and then $\epsilon$ decreases linearly during 600 training steps.
During the whole training process, the tuple $\langle s,u,r,s'\rangle$ of state transitions is collected from recent training episodes (up to 5000) and stored in the replay buffer, which is used for uniform sampling of batches of size $b=32$ to calculate the loss function~\eqref{loss-QMIX}. The weights of the target network are updated every 150 training episodes. For simplicity the agents' networks share the same weights. Additionally, the agent index $i$ is integrated into each observation vector to introduce the training versatility~\cite{rashid2020monotonic}. The discount factor in Eq.~\eqref{y-tot-QMIX} is set to $\gamma=0.99$. For training the network weights, the ADAM optimizer is used~\cite{kingma2014adam} with the learning rate set at $\alpha=10^{-4}$.

%\subsection{Simulations setup}\label{subsect:setup}

\subsection{Combinatorial optimization}\label{subsect:combinatorial}

In this subsection we consider Max-Cut combinatorial optimization problem. Each problem instance in Max-Cut is represented as a graph $G=(V,E)$ determined by the sets of its vertices $V$ and edges $E$. The Max-Cut problem can be mapped into a standard Hamiltonian minimization problem~\cite{farhi2014quantum, crooks2018performance,zhou2020quantum} with
\begin{equation}\label{H-maxcut}
    H =  \frac{1}{2} \sum_{(i,j) \in E} (\sigma^z_i \sigma^z_j - I). 
\end{equation}
The QAOA~\cite{farhi2014quantum} is considered as a standard tool for solving combinatorial problems such as Max-Cut on a quantum computer. In QAOA, the expectation value~\eqref{cost} of the problem Hamiltonian~\eqref{H-maxcut} is minimized using the variational ansatz
\begin{equation}\label{QAOA-ansatz}
    \ket{\psi_p(\bm{\beta},\bm{\gamma})} 
    =  \prod_{k=1}^p \biggl[ e^{-i\beta_k H_x  } e^{-i \gamma_k H} \biggr] \ket{+}^{\otimes n},
\end{equation}
where $\ket{+} = (\ket{0} + \ket{1})/\sqrt{2}$ is the initial quantum state, $H_x = \sum_i \sigma^x_i$ is the mixer Hamiltonian, $p$ is the circuit depth, and $ \bm{\theta} \equiv (\bm{\beta},\bm{\gamma}) \in [0,\pi)^{p} \times [0,2\pi)^{p}$ are the variational parameters.
Note that the ansatz~\eqref{QAOA-ansatz} is problem-specific since it is constructed based on the problem Hamiltonian $H$.

In what follows, we apply the proposed MARL-QAS approach to design PQCs tailored for a certain class of Max-Cut problem instances, rather than for a single instance. Specifically, we aim for constructing circuits well-suited for any of Max-Cut instances from the class of 3-regular graphs. This idea is motivated by the structural similarity observed across Max-Cut problem instances for regular graphs. In fact, the optimal ansatz parameters in QAOA~\eqref{QAOA-ansatz} tend to converge to instance-independent values for most cases~\cite{wurtz2021fixed, galda2021transferability}.
Note that, in general, our MARL-QAS algorithm may effectively brute-force the solution for any specific instance. This can be done by constructing factorizable circuits composed of local operations from~\eqref{G-set-reduced}, for instance $R^x_j(\theta)$ rotations applied to each qubit $j$, that directly prepare the solution bit string. However, such circuits suffer from an extremely poor trainability since they provide a cost landscape~\eqref{cost} swamped with a large number of local minima~\cite{holmes2022connecting}. 

For the circuits designed by MARL-QAS for solving the Max-Cut problem~\eqref{H-maxcut}, we enforce a layer-wise-like structure inspired by the QAOA ansatz~\eqref{QAOA-ansatz}. On each environment step of the algorithm, we assume that a single common parameter is shared across all parameterized gates added by RL agents. This allows to reduce the total number of parameters and provides a more fair comparison to standard QAOA~\eqref{QAOA-ansatz}.

For each system size $n$, we take all non-isomorphic 3-regular graphs and split them into a training set ($M$ instances) and a test set ($K$ instances). Note that for each $n$ there exists only a limited number of non-isomorphic 3-regular graphs~\cite{wormald1999models}. Next, we apply our MARL-QAS method to design a PQC well-suited for minimizing all instances in the training set. This can be done by modifying the reward metric $\eta$ in Eq.~\eqref{reward} used in the MARL algorithm. We introduce the modified metric as an average over instances in the training set, 
\begin{equation}\label{MaxCut-reward}
    \eta = M^{-1} \sum_{m=1}^M \eta_m,
\end{equation}
where each $\eta_m$ is calculated based on the same ansatz, but using variational parameters optimized for each specific instance $m$. 
For systems of size $n$, we design between 5 and 25 various circuits depending on $n$ by running the MARL-QAS algorithm with different seeds. In these runs, we also vary the maximum allowed environment steps number, which can be viewed as a hyperparameter that controls the maximum depth of a designed circuit.

\begin{table*}[!ht]
\centering
\caption{Characteristics of quantum circuits designed by the MARL-QAS for solving Max-Cut problem instances~\eqref{H-maxcut} for the class of 3-regular graphs in comparison to the standard QAOA circuit~\eqref{QAOA-ansatz} for different system sizes $n$. For each $n$, $M$ 3-regular graphs are sampled in the training set and $K$ in the test set. The calculated characteristics include the approximation ratio $\eta$ averaged over all instances in the test set, the number of two-qubit gates $N_{2q}$ and the number of variational parameters $N_{\rm par}$. The QAOA circuit depth $p$ was chosen as the smallest depth, sufficient to reach the threshold of $\eta = 0.95$.}
\label{tab:MaxCut_main}
\begin{tabular}{@{} >{\centering\arraybackslash}m{1cm}  >{\centering\arraybackslash}m{1cm}  >{\centering\arraybackslash}m{1.5cm} *{6}{>{\centering\arraybackslash}m{2.2cm}} @{}}
\toprule
\multirow{2}{*}{\textbf{$n$}} &\multirow{2}{*}{\textbf{$M/K$}} &\multirow{2}{*}{\textbf{$p$}} & \multicolumn{2}{c}{$\eta$} & \multicolumn{2}{c}{$N_{2q}$} & \multicolumn{2}{c}{$N_{\rm par}$} \\
\cmidrule(lr){4-5} \cmidrule(lr){6-7} \cmidrule(lr){8-9}
&& & \textbf{QAOA} & \textbf{MARL-QAS} & \textbf{QAOA} & \textbf{MARL-QAS} & \textbf{QAOA} & \textbf{MARL-QAS} \\
\midrule
4&1/0 & 2 & 0.99 & 0.99 & 24 & 14 & 4 & 5 \\
6& 1/1 & 2 & 0.98 & 0.98 & 36 & 20 & 4 & 6 \\
8& 3/2& 3 & 0.98 & 0.99& 72 & 23 & 6 & 8 \\
10& 12/7& 4 & 0.97 & 0.97 & 120 & 40 & 8 & 12 \\
12 &70/15& 5 & 0.96 & 0.97 & 180 & 71 & 10 & 13 \\
\bottomrule
\end{tabular}
\end{table*}

Once the training is complete, we evaluate the performance of designed circuits on Max-Cut problem instances from the test set. We consider a designed circuit to be satisfactory, if the approximation ratio \eqref{approx-ratio} obtained using this circuit reaches or exceeds the threshold of $\eta = 0.95$ for all instances in both the training and test sets. For each $n$, we pick the best satisfactory circuit with the largest $\eta$ averaged over all instances in the test set and compare it to the QAOA circuit~\eqref{QAOA-ansatz}. The QAOA circuit depth required to reach the threshold for 3-regular graphs was chosen based on the results of~\cite{zhou2020quantum}, where the QAOA performance has been studied for Max-Cut on 3-regular graphs for up to $n=22$ qubits.

\begin{figure}[b]
    \centering
    \includegraphics[width=0.85\linewidth]{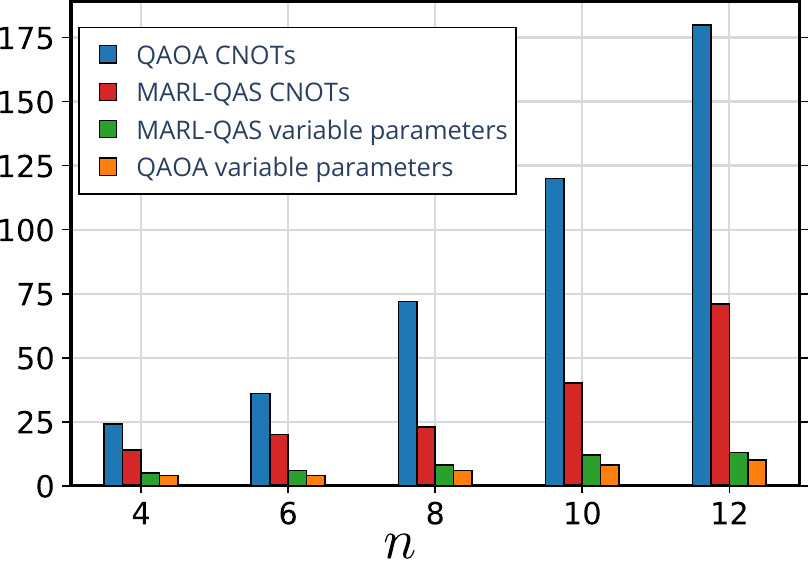}
    \caption{Number of two-qubit CNOT gates and number of variational parameters for circuits designed by the MARL-QAS algorithm for solving Max-Cut problem instances~\eqref{H-maxcut} for the class of 3-regular graphs in comparison to the standard QAOA circuit~\eqref{QAOA-ansatz} for different system sizes $n$. For each $n$, the minimal QAOA circuit depth $p$ is taken, which is required to reach the threshold of $\eta = 0.95$ in~\eqref{approx-ratio} (see Table~\ref{tab:MaxCut_main}).} 
    \label{fig:MaxCut-main}
\end{figure}

In Table~\ref{tab:MaxCut_main}, we compare the characteristics of the best circuit designed by MARL-QAS versus standard QAOA~\eqref{QAOA-ansatz} for each system size $n$ (see also Fig.~\ref{fig:MaxCut-main}). The performance of both circuits in terms of the approximation ratio~\eqref{approx-ratio} is approximately the same. The number of variational parameters used in the MARL-QAS circuits is slightly larger (13 vs 10 in QAOA for $n=12$). However, these circuits require about 2--3 times less number of CNOT gates as compared to QAOA, which favors their implementation on NISQ devices~\cite{Preskill2018quantumcomputingin}. To calculate the number of CNOT gates used in QAOA we note that an $n$-vertex 3-regular graph has $3n/2$ edges; for each edge, one has a $R_{zz}({\theta})$ two-qubit rotational gate, which requires two CNOTs for its implementation. The QAOA circuit~\eqref{QAOA-ansatz} of depth $p$ requires $p$ repetitions of these gates. Thus, the total number of CNOTs in QAOA for 3-regular graphs is equal to $3np$. 

\begin{figure}[b]
    \centering
    \includegraphics[width=0.9\linewidth]{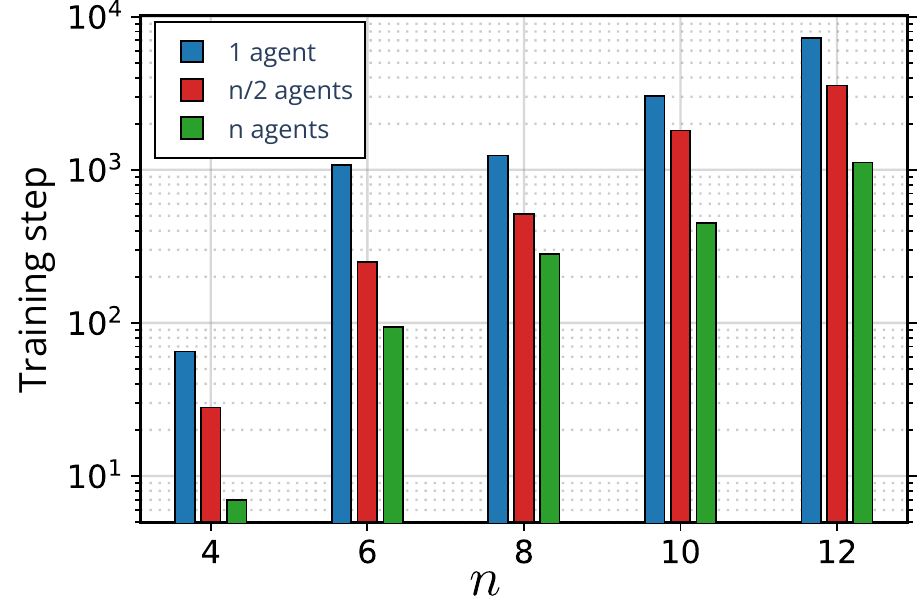}
    \caption{Number of training steps required to converge to the optimal circuit composition in the RL-based QAS with the approximation ratio $\eta \geq 0.95$ for Max-Cut problem instances~\eqref{H-maxcut} for the class of 3-regular graphs for different system sizes~$n$. 
    The results are averaged over multiple runs using the single-agent and MARL-QAS algorithms for $m=n/2$ and $n$ agents.} 
    \label{fig:MaxCut-steps}
\end{figure}

In Fig.~\ref{fig:MaxCut-steps}, we examine the effect of multiple agents on the training process in the RL-based QAS. We run our MARL-QAS algorithm upon varying the number of RL agents $m=1$, $n/2$ and $n$.
In each case the algorithm is able to design a satisfactory circuit. However, the number of training steps required to design such a circuit can be significantly reduced by increasing the number of RL agents. 
From Fig.~\ref{fig:MaxCut-steps} we see that the number of training steps in DQN/QMIX algorithm can be reduced by almost an order of magnitude using $n$ agents (with each agent supervising a single qubit). For instance, for $n=4$ we see that 72 training steps are required for convergence for a single-agent-based RL ($m=1$), while using $m=4$ agents in MARL needs only 7 steps. This advantage of MARL preserves for larger system sizes. For $n=12$, we observe {$7.2 \cdot 10^3$ training steps required for a single agent against $1.1 \cdot 10^3$ for $m=12$ agents. 
% Thus, the MARL approach accelerates the convergence to the optimal composition of a PQC in QAS.
Specifically, each training step in RL consists of optimizing the weights of the QMIX network~\eqref{loss-QMIX} as well as multiple environment steps, which require the optimization of a PQC~\eqref{cost} for calculating the reward~\eqref{reward}. 
Therefore, the overall computational cost for the RL-based QAS is significantly reduced in MARL since it requires less number of evaluations of the cost functions~\eqref{loss-QMIX} and \eqref{cost}. 
The latter is especially important for the quantum part of the algorithm, because each evaluation of the expectation value~\eqref{cost} in VQA is costly. 

\FloatBarrier

In Fig.~\ref{fig:MaxCut-ansatz}, we demonstrate an example of the ansatz designed by the MARL-QAS algorithm for $n=8$ qubits.
The designed ansatz prepares the bit string solution using not only single qubit operations: It exhibits a non-trivial entanglement structure and can not be factorized into smaller circuits. The ansatz makes use of a ladder of CNOTs, commonly encountered in variational quantum ansaetze, while also maintaining correlations between circuit parameters, reminiscent of QAOA. Yet, tailored to a specific problem class, the circuit requires fewer layers to achieve the target threshold when minimizing $3$-regular graphs.

\begin{figure}[ht]
    \centering 
    \includegraphics[width=\linewidth]{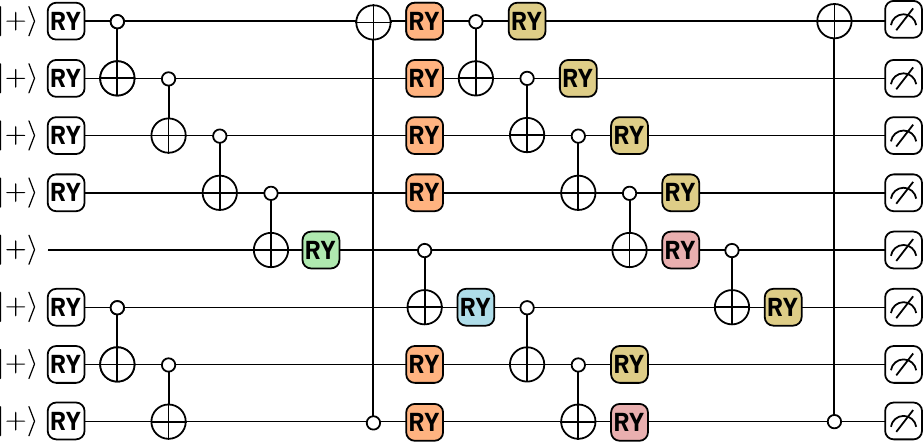}
    \caption{Example of the circuit designed by the MARL-QAS algorithm for solving the Max-Cut problem instances~\eqref{H-maxcut} for the class of 3-regular graphs for the system size $n=10$. Single qubit gates with the same color correspond to rotation by the same variable parameter.} 
    \label{fig:MaxCut-ansatz}
\end{figure}

\subsection{Results for Schwinger model}\label{subsect:Schwinger}

Next, we benchmark the proposed MARL-QAS algorithm on the problem of ground state energy search for Schwinger model Hamiltonian~\cite{Kokail2019}. The latter is a prominent example of a model with a strongly entangled ground state. The problem Hamiltonian reads
\begin{equation}\label{H-Schwinger}
    H \! = \! w \! \sum_{j=1}^{n-1} \! \left[ \sigma_j^+  \! \sigma_{j+1}^- \! \!+ \! \text{H.c.}  \right] 
    + \frac{m_0}{2}\!\sum_{j=1}^{n} (-1)^j  \sigma^z_j
    + \bar{g} \!\sum_{j=1}^{n}\! L_j^2,
\end{equation}
where $\sigma^{\pm}_j = \frac12 (\sigma^x_j \pm i\,\sigma^y_j)$, $L_j = \epsilon_0 - \frac{1}{2} \sum_{\ell = 1}^{j} ( \sigma^z_\ell + (-1)^\ell)$, $w$ is the spin flip-flop coupling, $m_0$ is the bare mass, $\bar{g}$ is the electric field coupling, and $\epsilon_0$ is the background electric field. In our simulations, in Eq.~\eqref{H-Schwinger} we set $w = 1$ as the energy unit, $m_0=1$, $\bar{g} = 1$ and $\epsilon_0 = 0$.

A standard approach for minimizing~\eqref{H-Schwinger} is to apply VQE~\cite{tilly2022variational} using a well-known expressive ansatz, for instance, the hardware-efficient ansatz (HEA)~\cite{Kandala2017}. This ansatz consists of $L$ layers: each comprises single-qubit rotations $R^\alpha(\theta)$, $\alpha=x,y,z$ applied to each qubit, followed by CNOT gates arranged in the ladder pattern,
\begin{eqnarray}
    U(\bm{\theta}) & = & \prod_{k=1}^{L} \, \biggl[ \, \prod_{j=1}^{n-1} \mathrm{CNOT}_{j,j+1} \nonumber \\
    &\times& \prod_{j=1}^n R^z_j(\theta_{k,j,3}) R^y_j(\theta_{k,j,2})  R^x_j(\theta_{k,j,1} ) \biggr]. \label{HEA}
\end{eqnarray}
This ansatz belongs to the search space of our QAS algorithm since it is composed from the same set~\eqref{G-set-reduced} of rotational and entangling gates. HEA~\eqref{HEA} is known to be highly expressive, but suffers from trainability limitations such as barren plateaus~\cite{holmes2022connecting}. However, by designing a problem-specific quantum circuit, one can identify an ansatz that overcomes these limitations. 

For each system size $n$, we use the proposed MARL-QAS approach to design an ansatz well-suited for minimizing the Schwinger Hamiltonian~\eqref{H-Schwinger}. Similarly to the case of combinatorial problems in Section~\ref{subsect:combinatorial}, we perform from 5 to 25 runs of MARL-QAS with different seeds depending on the problem size $n$. After that, we pick the best circuit in terms of the approximation ratio~$\eta$. 
The performance of the designed circuits in VQE is compared to HEA~\eqref{HEA}. The depth $L$ of HEA is ranging from $3$ to $12$ layers depending on the number of qubits $n$ (the first two columns in Table~\ref{tab:schwinger_main}). For minimizing the Schwinger model~\eqref{H-Schwinger} with HEA~\eqref{HEA}, we perform up to 200 iterations of the ADAM optimizer with the learning rate $\alpha=0.1$. In attempt to find the global minimum of the variational parameters for HEA, we pick the best result from $3$ runs of the optimizer starting from random initial parameters.

%The training of this ansatz was configured such that for each number of qubits the number of used layers varied from 3 to 12. In addition, for each case the optimization was performed with 3 different seeds each having random parameter initialization. ADAM algorithm was chosen as the gradient optimizer with learning rate set to $\alpha=0.1$. 
\begin{table*}[!t]
\centering
\caption{Characteristics of quantum circuits designed by the MARL-QAS for minimizing the Schwinger Hamiltonian~\eqref{H-Schwinger} in comparison to the standard HEA~\eqref{HEA} for different system sizes $n$. The calculated characteristics include the approximation ratio $\eta$, the number of two-qubit gates $N_{2q}$ and the number of variational parameters $N_{\rm par}$. For each $n$, the minimal depth of HEA $L$ was chosen as the smallest depth, sufficient to reach the threshold of $\eta = 0.95$.}
\label{tab:schwinger_main}
\begin{tabular}{@{} >{\centering\arraybackslash}m{1cm}>{\centering\arraybackslash}m{1.5cm} *{6}{>{\centering\arraybackslash}m{2.2cm}} @{}}
\toprule
\multirow{2}{*}{$n$} & \multirow{2}{*}{$L$} & \multicolumn{2}{c}{$\eta$} & \multicolumn{2}{c}{$N_{2q}$} & \multicolumn{2}{c}{$N_{\rm par}$} \\
\cmidrule(lr){3-4} \cmidrule(lr){5-6} \cmidrule(lr){7-8} &
 & \textbf{HEA} & \textbf{MARL-QAS} & \textbf{HEA} & \textbf{MARL-QAS} & \textbf{HEA} & \textbf{MARL-QAS} \\
\midrule
4 &3& 0.99  & 0.97 & 12 & 5 & 36 & 9 \\
6 &5& 0.97  & 0.97 & 30 & 27 & 90 & 55 \\
8 &6&  0.96 & 0.98 & 48 & 34 & 144 & 71 \\
10 &8&  0.98& 0.97 & 80 & 41 & 240 & 82 \\
12 &10& 0.97 & 0.97 & 120 & 56 & 360 & 108 \\
\bottomrule
\end{tabular}
\end{table*}

\begin{figure}[b]
    \centering
    \includegraphics[width=0.85\linewidth]{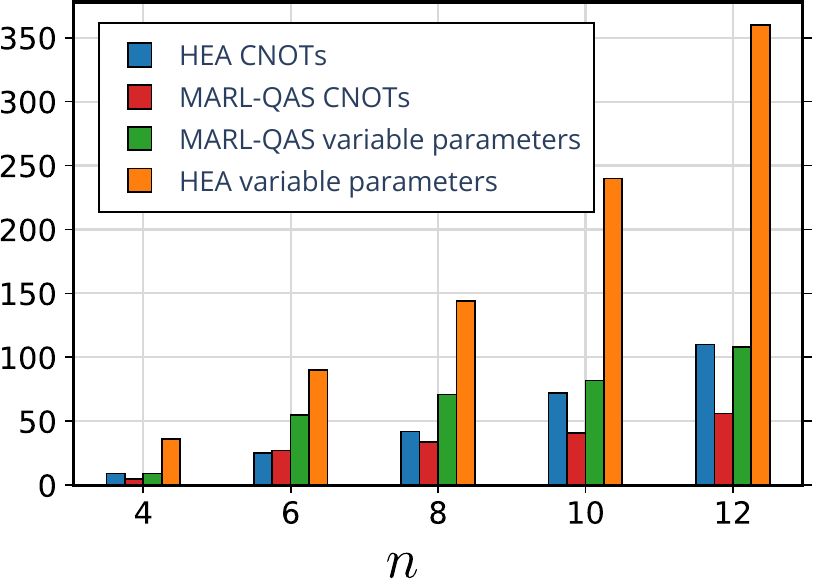}
    \caption{Number of two-qubit CNOT gates and number of variational parameters for circuits designed by the MARL-QAS algorithm for minimizing the Schwinger Hamiltonian~\eqref{H-Schwinger} in comparison to the standard HEA~\eqref{HEA} for different system sizes $n$. For each $n$, the minimal depth of HEA $L$ was chosen as the smallest depth, sufficient to reach the threshold of $\eta = 0.95$ in~\eqref{approx-ratio}  (see Table~\ref{tab:schwinger_main}).} 
    \label{fig:Schwinger-main}
\end{figure}

In Table~\ref{tab:schwinger_main}, we compare the properties of the circuits designed by MARL-QAS to HEA~\eqref{HEA} for each system size $n$, graphically represented at Fig.~\ref{fig:Schwinger-main}. Their performance in terms of the approximation ratio~\eqref{approx-ratio} is almost the same. However, the MARL-QAS circuits require about half as much two-qubit gates as in HEA. Moreover, these circuits provide another significant advantage over HEA as they are parameterized by a drastically smaller number of parameters. For instance, at $n=4$ we use HEA~\eqref{HEA} of depth $L=3$ with 36 variational parameters, while the designed circuit has only 9 parameters. This advantage persists with the system size: at $n=12$ qubits and $L=10$ layers HEA has 360 parameters in comparison to only 108 parameters in the MARL-QAS circuit. Thus, our algorithm allows to design PQCs more favorable for the optimization as compared to the standard circuits, while keeping the designed circuit as shallow as possible.

%For the VQE problem, the properties of the ansatze produced by the MARL-QAS algorithm, as well as the corresponding parameters of HEA for the same problem are presented in the Table.\ref{tab:schwinger_main}. Number of single-qubit gates is muted in this table as well for the same reasons as for the MaxCut problem. However, in terms of the number of two-qubit gates and variable parameters, the difference between the MARL-QAS algorithm and the application of regular HEA ansatz is not that dramatic as for the MaxCut problem. For the most part this is due to the fact that for the MARL-QAS algorithm the number of variable parameters was the same as the total amount of variational parameters (in other words, all parameters are independent). Making parameters independent was necessary for this problem since for the layer-like structure with shared variable parameters the total amount of one-qubit rotational gates was several times more than for HEA. Consequently, the only advantage that MARL-QAS algorithm could offer was the nuanced placement of gates so that the inherent properties of Schwinger Hamiltonian could be addressed more accurately. Still, the algorithm was able to provide more efficient ansatz in comparison with HEA.

\begin{figure}[b]
    \centering
    \includegraphics[width=\linewidth]{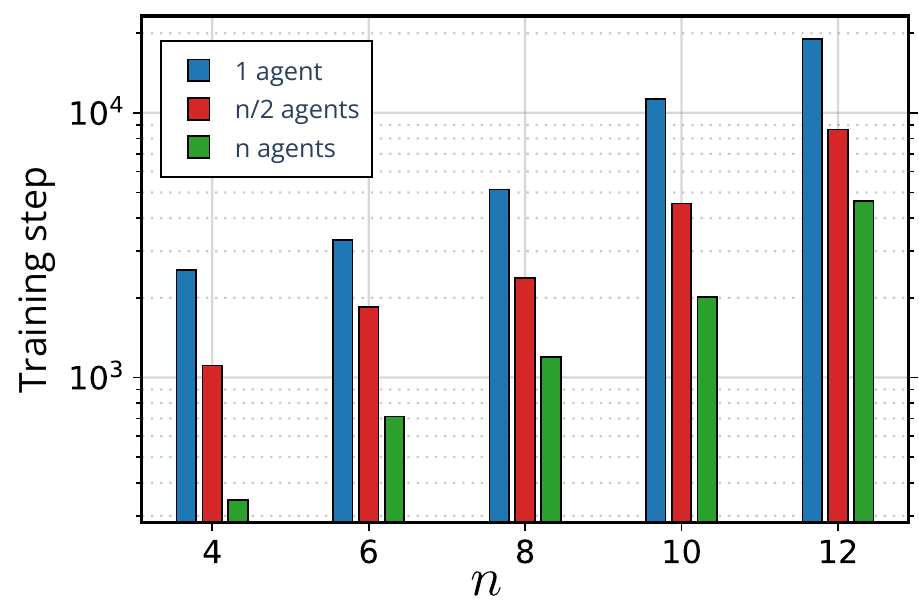}
    \caption{Number of training steps required to converge to the optimal circuit composition in the RL-based QAS with the approximation ratio $\eta \geq 0.95$ for minimizing the Schwinger Hamiltonian~\eqref{H-Schwinger} for different system sizes~$n$. 
The results are averaged over multiple runs using the single-agent and MARL-QAS algorithms for $m=n/2$ and $n$ agents.} 
    \label{fig:Schwinger-steps}
\end{figure}

Similarly to the Max-Cut problem in Section ~\ref{subsect:combinatorial}, in Fig.~\ref{fig:Schwinger-steps} we demonstrate the effect of multiple RL agents in the RL-based QAS on the number of training steps for searching a problem-specific PQC for the Schwinger model~\eqref{H-Schwinger}. As one can see the results are qualitatively the same as for the combinatorial optimization shown in Fig.~\ref{fig:MaxCut-steps}. Quantitatively, we observe that the number of training steps required to design circuits for the Schwinger model~\eqref{H-Schwinger} is slightly larger. This is due to the fact that it is harder to prepare an entangled ground state of~\eqref{H-Schwinger} as compared to bit strings in combinatorial optimization problems such as~\eqref{H-maxcut}. However, the advantage of using multiple agents becomes even more pronounced for the Schwinger model. For instance, for $n=4$ qubits one needs to perform $10^3$ training steps in the single-agent RL to design a satisfactory circuit, while it takes only about $200$ steps for the same process using $m=4$ agents in MARL-QAS. For $n=12$, we have about $10^4$ training steps in the single-agent-based RL algorithm in comparison to $2 \cdot 10^3$ steps for $m=12$ agents in MARL. Overall, for all considered problem sizes, we observe approximately a fivefold reduction in the number of training steps in MARL-QAS compared to the single agent RL approach.

\FloatBarrier

\section{Conclusions}\label{sect:conclusion}

In the framework of quantum architecture search (QAS) one of the main obstacles is the exploration of huge search space of possible circuit realizations. One of the possible solutions is to automatically design quantum circuits by means of Markov decision processes, where gates are added sequentially. This can be modeled with the classical reinforcement learning (RL). The approach is based on the idea of circuit construction through a sequence of actions performed by an agent with a trainable policy.
A few existing algorithms are based on a single-agent RL, which has a limited room for a scalability improvement. 

In this paper, we proposed a Multi-Agent Reinforcement Learning approach for Quantum Architecture Search (MARL-QAS). This approach trains multiple agents in cooperative manner to design a quantum circuit, tailored to specific problems, from a prescribed set of gates. For training a cooperative multi-agent system, we used one of the most prominent classical QMIX algorithm. We benchmarked the developed algorithm on combinatorial optimization (Max-Cut) and ground state search (Schwinger Hamiltonian) problems. When compared to conventional solutions using quantum approximate optimization (QAOA) and hardware efficient ansaetze (HEA) with the same approximation ratio, the designed circuits exhibit a substantially lower number of entangling gates. This can be crucial for implementing the circuits on real quantum devices. The MARL-QAS circuits designed for the Schwinger Hamiltonian are also parameterized with a smaller number of parameters as compared to the standard HEA. The latter makes our circuits more favorable for optimization.

%The MARL-QAS approach becomes especially efficient in designing dense quantum circuits, i.e., each qubit is a subject to approximately the same number of quantum gates.
% In this paper, a novel approach for distributed quantum architecture search based on the multi-agent reinforcement learning is proposed. 
%For the tasks of ansaetze states reconstruction, the algorithm has produced circuits with lesser amount of entangling gates which output states with high fidelity.
%Furthermore, the algorithm has achieved better performance in comparison with one of the most prominent quantum machine learning algorithms, QAOA. 

We demonstrated that the multi-agent approach facilitates faster and more effective exploration compared to the single-agent RL.
Specifically, we showed that by using multiple RL agents, one can significantly reduce the number of required training steps in the RL algorithm. Since each training step requires several evaluations of the reward based on optimizing a quantum circuit, MARL-QAS thus allows to reduce the overall number of calls to a quantum computer. The achieved convergence acceleration can be explained as follows. For variational quantum circuits, where a similar number of gates are applied to each qubit, multi-agent algorithms prove to be statistically efficient by 
simultaneously adding gates to each wire in the circuit. This is more efficient than for the single-agent RL, where on each environment step only a single gate is added to the circuit. 

Note that a similar idea could be implemented using the RL action branching method \cite{tavakoli2018action}. This method decomposes the action space into several branches, where one can apply actions simultaneously for each action branch. However, this method still requires the centralized shared decision module. This crucial detail makes the action branching method a less promising candidate for distributed execution, while the multi-agent approach naturally fits the recent development of distributed quantum hardware. In particular, once the training of MARL-QAS is complete, the agents can perform actions independently according to their own optimal policies. Thus, MARL-QAS can be applied to design circuits for the distributed QAS framework, where a quantum system is partitioned between multiple small quantum processors, each supervised by its own RL agent. 

%The proposed MARL algorithm can be further expanded and improved. %For example, upon training on a real hardware, the computational overhead of simulating the entire system will be neglected, further enhancing the advantages of the algorithm. 
%Furthermore, overall training approach can be extended to meta-RL scenarios where agents can be trained to produce efficient quantum circuits for solving numerous quantum machine learning tasks. Additionally, current approach can be improved as well by additional hyperparameters tweaking, addition of other quantum instructions  and modification of the training loop. 

% the case could be extended to shots mode as well. nobody cares

%\begin{acknowledgments}

%The authors acknowledge the usage of the computational resources at the Skoltech supercomputer ``Zhores''~\cite{Zhores}. 

%\end{acknowledgments}

\section*{Data Availability Statement}

The data and code that support the findings of
this study is publicly available at \href{https://github.com/WolvenAnthros/dqas-marl.git}{Github}. 

%\nocite{*}
\bibliography{refs.bib}
\bibliographystyle{unsrt}

\appendix

\end{document}